\documentclass[10pt,conference]{IEEEtran}
\IEEEoverridecommandlockouts                            
\usepackage{cite,graphicx,amsmath,amssymb}
\usepackage{subfigure}
\usepackage{citesort}
\usepackage{fancyhdr}
\usepackage{dsfont}
\usepackage{array,color}
\usepackage{amsfonts}
\usepackage{hhline}
\usepackage{cases}
\newcommand{\dis}{\stackrel{d}{\sim}}

\newtheorem{theorem}{Theorem}

\newtheorem{lemma}{Lemma}

\newtheorem{corollary}{Corollary}

\newtheorem{proposition}{Proposition}

\newcommand{\define}{\stackrel{\Delta}{=}}

\begin{document}
%

\title{Analysis and Optimization of Interference Nulling in Downlink Multi-Antenna HetNets with Offloading}
\author{
Yueping Wu$^{*}$, \, Ying Cui$^{\ddag}$, \,
Bruno Clerckx$^{*\dagger}$\\
\begin{minipage}{2\columnwidth}
\begin{center}
\small $^{*}$Department of Electrical and Electronic Engineering, Imperial College London\hspace{4mm}$^{\dagger}$School of Electrical Engineering, Korea University\\
$^{\ddag}$Department of Electronic Engineering, Shanghai Jiao Tong University\\
\end{center}
\end{minipage}
\thanks{This work was partially supported by the Seventh Framework Programme
for Research of the European Commission under grant number HARP-318489.}
}



\maketitle

\begin{abstract}
Heterogeneous networks (HetNets) with offloading is considered as an effective way to meet the high data rate demand of future wireless service. However, the offloaded users suffer from strong inter-tier interference, which reduces the benefits of offloading and is one of the main limiting factors of the system performance. In this paper, we investigate the use of an interference nulling (IN) beamforming scheme to improve the system performance by carefully managing the inter-tier interference to the offloaded users in downlink two-tier HetNets with multi-antenna base stations. Utilizing tools from stochastic geometry, we derive a tractable expression for the rate coverage probability of the IN scheme. Then, we optimize the design parameter, i.e., the degrees of freedom that can be used for IN, to maximize the rate coverage probability. Specifically, in the asymptotic scenario where the rate threshold is small, by studying the order behavior of the rate coverage probability, we characterize the optimal design parameter. For the general scenario, we show some properties of the optimal design parameter. Finally, by numerical simulations, we show the IN scheme can outperform both the simple offloading scheme without interference management and the almost blank subframes scheme in 3GPP LTE, especially in large antenna regime. 
\end{abstract}

\IEEEpeerreviewmaketitle

\section{Introduction}\label{sec:intro}
The next generation of wireless networks will see a significant increase in the number of wireless users and the scope of high data rate applications, leading to a high data rate demand. The conventional cellular solution, which comprises of high power base stations (BSs) covering a large cellular area, will not be able to scale with the increasing data rate demand of these future networks. A promising solution is the deployment of low power small cell nodes overlaid with high power macro-BSs, so called heterogeneous networks (HetNets), which are capable of aggressive reuse of existing spectrum assets, and thus lead to the support of future high data rate applications. 

Due to the larger power at the macro-BSs, more users intend to associate with the macro-cell tier, which causes the problem of \emph{load imbalancing} between the macro-cell tier and the small-cell tier. Load imbalancing is recognized as one of the main bottlenecks for the performance of HetNets \cite{jo12,singh13may}. As such, to remit this problem, it is suggested to offload users from the heavily loaded macro-cell tier to the lightly loaded small-cell tier. Under the offloading scheme, the nearest macro-BS of each offloaded user, which provides the strongest signal among all the BSs, now becomes the dominant interferer of this offloaded user. Thus, after offloading, the offloaded users have degraded signal-to-interference ratio (SIR). The performance of the offloaded users is one of the limiting factors of the network performance. Several interference management techniques have been considered to improve the performance of the offloaded users in HetNets with offloading. One such technique is almost blank subframes (ABS) in 3GPP LTE. In ABS, (time or frequency) resource is partitioned, whereby the offloaded users and the other users are served using different portions of the resource. The performance of ABS in HetNets with offloading was analyzed in \cite{singh13} using tools from stochastic geometry. Another interference management technique was proposed in \cite{sakr14} to reduce the interference to offloaded users, where each offloaded user is jointly served by both of its nearest macro-BS and nearest pico-BS. The performance of this joint transmission scheme in large HetNets was analyzed using similar tools as in \cite{singh13}. Note that these works only considered HetNets with single-antenna BSs. 

Deploying multiple antennas in HetNets is another solution to provide high data rates for future wireless service. Futhermore, with multiple antennas, more effective interference management techniques (e.g., coordinated beamforming \cite{cui12}) can be implemented. For example, a HetNet with a single multi-antenna macro-BS was investigated in \cite{Hosseini13,Adhikary14}, where the multiple antennas at the macro-BS are used for serving its scheduled users as well as mitigating interference to the receivers in small cells using different interference coordination schemes. These schemes have been analyzed and shown to have performance improvement. However, since only one macro-BS is considered, the results obtained in \cite{Hosseini13,Adhikary14} can not reflect the macro-tier interference, and thus can not offer accurate insights for practical networks. In \cite{xia13}, interference coordination among a \emph{fixed} number of neighboring BSs was investigated in downlink large multi-antenna HetNets. However, this scheme may not fully exploit the spatial properties of the interference in large HetNets. Moreover, offloading was not considered in \cite{xia13}. So far, it is still not clear how the interference coordination schemes and the system parameters affect the performance of large multi-antenna HetNets with offloading.


In this paper, we consider offloading in downlink two-tier large stochastic HetNets with multi-antenna BSs, and investigate the use of interference nulling (IN) beamforming to improve the performance of the offloaded users. The scheme has a design parameter, which is the degree of freedom $U$ that can be used at each macro-BS for IN to its offloaded users. In particular, each macro-BS utilizes the low-complexity zero-forcing beamforming (ZFBF) precoder to suppress interference to at most $U$ offloaded users as well as boost the desired signal to its scheduled user. Note that interference coordination using beamforming technique in large stochastic HetNets causes spatial dependency among macro and pico cells, which is quite difficult to analyze in general \cite{Adhikary14}. In this paper, by adopting appropriate approximations and utilizing tools from stochastic geometry, we first present a tractable expression for the rate coverage probability of the IN scheme. To our best knowledge, this is the first work analyzing the interference coordination technique in large stochastic multi-antenna HetNets with offloading. To further improve the rate coverage probability, we consider the optimization of the design parameter. Note that optimization problems in large HetNets with single-antenna BSs were investigated (see e.g., \cite{lin14}). In \cite{lin14}, the objective function is relatively simple, and bounds of the objective function are utilized to obtain near-optimal solutions. The optimization problem in large multi-antenna HetNets we consider is an integer programming problem with a very complicated objective function. Hence, it is quite challenging to obtain the optimal solution. First, for the asymptotic scenario where the rate threshold is small, by studying the order behavior of the rate coverage probability, we prove that the optimal design parameter converges to a fixed value, which is only related to the antenna number difference between each macro-BS and each pico-BS. Next, for the general scenario, we show that the optimal design parameter also depends on other system parameters. Finally, by numerical simulations, we show the IN scheme can outperform both the simple offloading scheme without interference management and ABS in 3GPP LTE, especially in large antenna regime.

\section{System Model}
\subsection{Downlink Two-Tier Heterogeneous Networks}
We consider a two-tier HetNet where a macro-cell tier is overlaid with a pico-cell tier, as shown in Fig.\ \ref{fig:HetNets}. The locations of the macro-cell BSs and the pico-cell BSs are spatially distributed as two independent Homogeneous Poisson point processes (PPPs) $\Phi_{1}$ and $\Phi_{2}$ with densities $\lambda_{1}$ and $\lambda_{2}$, respectively. The locations of the users are also distributed as an independent homogeneous PPP $\Phi_{u}$ with density $\lambda_{u}$. Without loss of generality (w.l.o.g.), denote the macro-cell tier as the $1$st tier and the pico-cell tier as the $2$nd tier. We consider the downlink transmission. Each macro-BS has $N_{1}$ antennas with a total transmission power $P_{1}$, each pico-BS has $N_{2}$ antennas with a total transmission power $P_{2}$ ($<P_{1}$), and each user is equipped with a single antenna. We consider both large-scale fading and small-scale fading. Specifically, due to large-scale fading, in the $j$th tier ($j=1,2$), transmitted signals with distance $r$ are attenuated by a factor $\frac{1}{r^{\alpha_{j}}}$, where $\alpha_{j}>2$ is the path loss exponent of the $j$th tier. For small-scale fading, we assume Rayleigh fading channels, i.e., each element of channel vectors is distributed as $\mathcal{CN}(0,1)$. 

\subsection{User Association}
We assume open access \cite{jo12}. Due to the larger power at the macro-BSs, the load imbalancing problem arises if the user association is according to the long-term average received power (RP). To remit the load imbalancing problem, the bias factor $B_{j}$ ($j=1,2$) is introduced to each tier, where $B_{2}>B_{1}$, to offload users from the macro-cell tier to the pico-cell tier. Specifically, user $i$ (denoted as $u_{i}$) is associated with the BS which provides the maximum \emph{long-term averaged} biased-received-power (BRP) (among all the macro-BSs and pico-BSs) at the user. Here, the long-term averaged BRP is defined as the average RP multiplied by a bias factor. This associated BS is called the \emph{serving BS} of user $i$. Note that within each tier, the nearest BS to user $i$ provides the strongest long-term averaged BRP in this tier. User $i$ is thus associated with the nearest BS in the $j^{*}_{i}$th tier if\footnote{In the cell selection procedure, the first antenna is normally used to transmit signal (using the total transmission power of each BS) for BRP determination according to LTE standards.} 
$j_{i}^{*}=  {\arg\:\max}_{j\in\{1,2\}}P_{j}B_{j}Z_{i,j}^{-\alpha_{j}}$, 
where $Z_{i,j}$ is the distance between user $i$ and its nearest BS in the $j$th tier. From the criterion described above, we observe that, for given $\{P_{j}\}$, $\{Z_{i,j}\}$ and $\{\alpha_{j}\}$, the user association is only affected by the bias factor ratio. Thus, w.l.o.g., we assume $B_{1}=1$ and $B_{2}=B>1$. After user association, each BS schedules its associated users according to TDMA, i.e., scheduling one user in each time slot, so that there is no intra-cell interference. 

\begin{figure}[t]
\centering
\subfigure[System Model ($U=1$)]{
\includegraphics[width=0.66\columnwidth]
{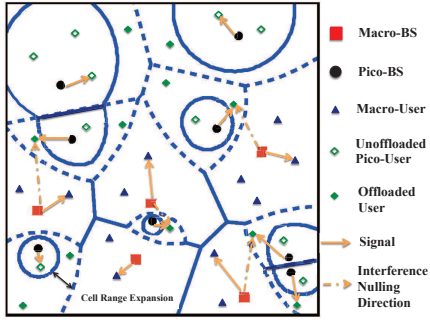}
\label{fig:HetNets}
}
\subfigure[User Set Illustration]{
\includegraphics[width=0.4\columnwidth]{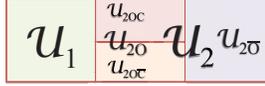}
\label{fig:user_set}
}
\caption{System model ($U=1$) and user set illustration.}\label{fig:model}
\vspace{-5mm}
\end{figure}

According to the above mentioned user association policy and the offloading strategy, as illustrated in Fig.\ \ref{fig:user_set}, all the users can be partitioned into the following three disjoint user sets: i) set of \emph{macro-users}: $\mathcal{U}_{1}=\left\{u_{i}|P_{1}Z_{i,1}^{-\alpha_{1}}\ge BP_{2}Z_{i,2}^{-\alpha_{2}}\right\}$, ii) set of \emph{unoffloaded pico-users}: $\mathcal{U}_{2\bar{O}}=\left\{u_{i}|P_{2}Z_{i,2}^{-\alpha_{2}}>P_{1}Z_{i,1}^{-\alpha_{1}}\right\}$, iii) set of \emph{offloaded users}: $\mathcal{U}_{2O}=\left\{u_{i}|P_{2}Z_{i,2}^{-\alpha_{2}}\le P_{1}Z_{i,1}^{-\alpha_{1}}<BP_{2} Z_{i,2}^{-\alpha_{2}}\right\}$,
where the macro-users are associated with the maco-BSs, the unoffloaded pico-users are associated with the pico-BSs (even without bias), and the offloaded users are offloaded from the macro-BSs to the pico-BSs (due to bias $B>1$). Moreover, $\mathcal{U}_{2}=\mathcal{U}_{2\bar{O}}\bigcup\mathcal{U}_{2O}$ is \emph{the set of pico-users}. 

\subsection{Performance Metric}
In this paper, we study the performance of the typical user denoted as\footnote{The index of the typical user and its serving BS is denoted as $0$.} $u_{0}$, which is located at the origin and is scheduled. 
Since HetNets are interference-limited, in this paper, we ignore the thermal noise in the analysis. Note that the analysis including thermal noise can be obtained in a similar way. Let $R_{0}=\frac{W}{L_{0}}\log_{2}\left(1+{\rm SIR}_{0}\right)$ denote the rate of the typical user, where $W$ is the available resource (e.g., time or frequency), $L_{0}$ is the total number of associated users (i.e., \emph{load}) of the typical user's serving BS, and ${\rm SIR}_{0}$ is the SIR of the typical user. We investigate the \emph{rate coverage probability} of the typical user, which is defined as the probability that the rate of the typical user is larger than a threshold \cite{singh13}, i.e.,
\begin{align}\label{eq:CPrate_def}
\mathcal{R}(\tau)&\define{\rm Pr}\left(R_{0}>\tau\right)={\rm Pr}\left(\frac{W}{L_{0}}\log_{2}\left(1+{\rm SIR}_{0}\right)>\tau\right)
\end{align}
where $\tau$ is the rate threshold. The rate coverage probability is a desired performance metric for applications with strict rate requirements, e.g., video services \cite{singh13may}.  

\section{Inter-tier Interference Nulling}\label{sec:ITIC_model}
In HetNets with offloading, the offloaded users (to the pico-cell tier) normally suffer from stronger interference than the macro-users and unoffloaded pico-users. \footnote{For each offloaded user, its nearest macro-BS, which provides the strongest long-term averaged RP, now becomes the dominant interferer of this offloaded user. However, for each macro-user or unoffloaded pico-user, the BS which provides the strongest long-term averaged RP is its serving BS. Therefore, the offloaded users suffer from the strongest interference.} The dominant interference to each offloaded user, which is caused by its nearest macro-BS, is one of the limiting factors of the system performance. In this section, we first investigate an inter-tier IN scheme to improve the performance of these offloaded users. Then, we obtain the SIR under the IN scheme. 

\subsection{Strategy Description}\label{sub:IN_intro}
We consider an inter-tier IN scheme to mitigate the dominant interference to the offloaded users. We first partition the offloaded users $\mathcal{U}_{2O}$ into two sets. Note that all the offloaded users may not be scheduled by their nearest pico-BSs simultaneously, as each BS schedules one user in each time slot. We refer to the offloaded users, which are offloaded from a macro-BS and are scheduled by their nearest pico-BSs in a particular time slot, as \emph{active offloaded users} of this macro-BS. In this scheme, each macro-BS conducts IN to some of its active offloaded users in a particular time slot, which are referred to as \emph{IN offloaded users} of this macro-BS. We refer to the remaining offloaded users of each macro-BS as \emph{non-IN offloaded users} of this macro-BS. Under the IN scheme, in a particular time slot, as illustrated in Fig.\ \ref{fig:user_set}, the offloaded users $\mathcal{U}_{2O}$ can be divided into two sets, i.e., $\mathcal{U}_{2O}=\mathcal{U}_{2OC}\bigcup \mathcal{U}_{2O\bar{C}}$, where $\mathcal{U}_{2OC}$ is the set of IN offloaded users and $\mathcal{U}_{2O\bar{C}}$ is the set of non-IN offloaded users. 

We now discuss how to determine the IN offloaded users of each macro-BS. In particular, each macro-BS, which has $N_{1}$ transmit antennas, makes use of at most $U$ ($U<N_{1}$) DoF to conduct IN. Note that $U$ is the design parameter of this scheme. When $U=0$, the IN scheme reduces to the one without interference management technique. Specifically, let $U_{2O_{a},\ell}$ denote the number of active offloaded users of macro-BS $\ell$, each of which is scheduled by a different pico-BS.  If $U_{2O_{a},\ell}\le U$, macro-BS $\ell$ performs IN to all of its $U_{2O_{a},\ell}$ active offloaded users using $U_{2O_{a},\ell}$ DoF. However, if $U_{2O_{a},\ell}>U$, macro-BS $\ell$ randomly selects $U$ out of $U_{2O_{a},\ell}$ active offloaded users according to the uniform distribution to perform IN using $U$ DoF. Hence, macro-BS $\ell$ performs IN to $u_{2OC,\ell}\define\min\left(U,U_{2O_{a},\ell}\right)$ out of $U_{2O_{a},\ell}$ active offloaded users. Note that the DoF used for IN (called IN DoF) at macro-BS $\ell$ is $u_{2OC,\ell}$. The remaining $N_{1}-u_{2OC,\ell}$ DoF at macro-BS $\ell$ is used for boosting the signal to its scheduled user. 

Now, we introduce the precoding vectors at macro-BSs and pico-BSs, respectively. First, each macro-BS utilizes the low-complexity ZFBF vector to serve its scheduled user and simultaneously perform IN to its IN offloaded users. Specifically,  denote $\mathbf{H}_{1,\ell}=\left[\mathbf{h}_{1,\ell}\; \mathbf{g}_{1,\ell1}\;\mathbf{g}_{1,\ell2}\;\ldots\;\mathbf{g}_{1,\ell u_{2OC,\ell}}\right]^{H}$, where\footnote{The notation $X \dis Y$ means that $X$ \emph{is distributed as} $Y$.} $\mathbf{h}_{1,\ell}\dis\mathcal{CN}_{N_{1},1}\left(\mathbf{0}_{N_{1}\times 1},\mathbf{I}_{N_{1}}\right)$ is the channel vector between macro-BS $\ell$ and its scheduled user, and $\mathbf{g}_{1,\ell i}\dis \mathcal{CN}_{N_{1},1}\left(\mathbf{0}_{N_{1}\times 1},\mathbf{I}_{N_{1}}\right)$ denotes the channel vector between macro-BS $\ell$ and its IN offloaded user $i$ $(i=1,\ldots,u_{2OC,\ell})$. The precoding matrix is designed to be $\mathbf{W}_{1,\ell}=\mathbf{H}_{1,\ell}^{H}\left(\mathbf{H}_{1,\ell}\mathbf{H}_{1,\ell}^{H}\right)^{-1}$. Then, the ZFBF vector at macro-BS $\ell$ is designed to be $\mathbf{f}_{1,\ell}=\frac{\mathbf{w}_{1,\ell}}{\|\mathbf{w}_{1,\ell}\|}$, where $\mathbf{w}_{1,\ell}$ is the first column of the precoding matrix $\mathbf{W}_{1,\ell}$. Next, the beamforming vector at each pico-BS is utilized to serve its scheduled user. Specifically, we use the maximum-ratio transmission scheme. The beamforming vector at pico-BS $\ell$ is $\mathbf{f}_{2,\ell}=\frac{\mathbf{h}_{2,\ell}}{\left\|\mathbf{h}_{2,\ell}\right\|}$, 
where $\mathbf{h}_{2,\ell}\dis\mathcal{CN}_{N_{2},1}\left(\mathbf{0}_{N_{2}\times 1},\mathbf{I}_{N_{2}}\right)$ is the channel vector between pico-BS $\ell$ and its scheduled user.

\begin{table}[t]
\caption{Parameter values when $u_{0}\in\mathcal{U}_{k}$ with $k\in\{1,2\bar{O},2OC,2O\bar{C}\}$}\label{tab:para_B2larger}
\begin{center}
\vspace{-3mm}
\begin{tabular}{|c!{\vrule width 1.5pt}c|c|c|c|}
\hline
$k$&$1$&$2\bar{O}$&$2OC$&$2O\bar{C}$\rule{0pt}{3mm}\\
\hhline{|=|=|=|=|=|}
$j_{k}$ &$1$&$2$ &$2$&$2$\\
\hline
$M_{k}$&$N_{1}-u_{2OC,0}$ & $N_{2}$&$N_{2}$&$N_{2}$\\
\hline
$\mathcal{B}_{k}$&$\{B_{1,0}\}$ & $\{B_{2,0}\}$&$\{B_{1,0},B_{2,0}\}$& $\{B_{2,0}\}$\\
\hline
\end{tabular}
\vspace{-6mm}
\end{center}
\end{table}

\subsection{SIR of the Typical User}
We now obtain the SIR expressions of the typical user $u_{0}\in \mathcal{U}_{k}$, for all $k\in\mathcal{K}\define\{1,2\bar{O},2OC,2O\bar{C}\}$. Specifically, the SIR of the typical user $u_{0}\in\mathcal{U}_{k}$ is given by\footnote{In this paper, all maco-BSs and pico-BSs are assumed to be active. The same assumption can also be seen in existing papers (see e.g., \cite{singh13,singh13may}).} ${\rm SIR}_{k,0} = \frac{\frac{\rho_{j_{k}}}{Y_{j_{k}}^{\alpha_{j_{k}}}}\left|\mathbf{h}_{j_{k},0}^{H}\mathbf{f}_{j_{k},0}\right|^{2}}{\sum_{j=1}^{2}\sum_{\ell\in\Phi\left(\lambda_{j}\right)\backslash \mathcal{B}_{k}}\frac{\rho_{j}\left|\mathbf{h}_{j,\ell0}^{H}\mathbf{f}_{j,\ell}\right|^{2}}{\left|D_{j,\ell0}\right|^{\alpha_{j}}}}$, 
where $\mathbf{h}_{j_{k},0}$ is the channel vector between $u_{0}\in\mathcal{U}_{k}$ and its serving BS, $\mathbf{f}_{j_{k},0}$ is the beamforming vector at the serving BS of $u_{0}\in\mathcal{U}_{k}$, with $\left|\mathbf{h}_{j_{k},0}^{H}\mathbf{f}_{j_{k},0}\right|^{2}\dis{\rm Gamma}\left(M_{k},1\right)$, $j_{k}$ and $M_{k}$ given by Table \ref{tab:para_B2larger}, $\mathbf{h}_{j,\ell0}$ is the channel vector between BS $\ell$ in the $j$th tier and $u_{0}$, $\mathbf{f}_{j,\ell}$ is the beamforming vector at BS $\ell$ in the $j$th tier, with $\left|\mathbf{h}_{j,\ell0}^{H}\mathbf{f}_{j,\ell}\right|^{2}\dis{\rm Gamma}(1,1)$ ($j=1,2$), $\rho_{j}$ is the transmit signal-to-noise ratio of the $j$th tier, $Y_{j_{k}}$ is the distance between $u_{0}\in\mathcal{U}_{k}$ and its serving BS, $\left|D_{j,\ell0}\right|$ is the distance between BS $\ell$ in the $j$th tier and $u_{0}$, and $\mathcal{B}_{k}$ (given by Table \ref{tab:para_B2larger}) is the BS set which does not cause interference to $u_{0}\in\mathcal{U}_{k}$. Here, $B_{j,0}$ is the nearest BS of $u_{0}$ in the $j$th tier. 



\section{Rate Coverage Probability Analysis}\label{sec:CPrate}
In this section, we investigate the rate coverage probability of the IN scheme. We notice that the rate coverage probability of the IN scheme is dependent on the probability mass functions (p.m.f.) of the active offloaded users $U_{2O_{a},0}$ of the serving macro-BS of $u_{0}$ when $u_{0}\in\mathcal{U}_{1}$ and the p.m.f. of the active offloaded users $\hat{U}_{2O_{a},0}$ of the nearest macro-BS of $u_{0}$ when $u_{0}\in\mathcal{U}_{2O}$. Note that the p.m.f.s of $U_{2O_{a},0}$ and $\hat{U}_{2O_{a},0}$ are related to Poisson distribution. But the exact distributions are unknown, as it is difficult to calculate due to the coupling between the active offloaded users and the pico-BSs. Here, we address this challenge by first deriving approximations for p.m.f.s of $U_{2O_{a},0}$ and $\hat{U}_{2O_{a},0}$. Based on the approximations and according to total probability theorem, we have the rate coverage probability $\mathcal{R}(\tau)$ as follows: \footnote{Due to page limit, we omit all the proofs. Please refer to \cite{wu14arxiv} for details.}
\begin{theorem}\label{cor:CPrate_overall}
The rate coverage probability is
\begin{align}\label{eq:CPrate_IN_overall}
\mathcal{R}(\tau)=&\mathcal{A}_{1}\mathcal{R}_{1}(\tau)+\mathcal{A}_{2\bar{O}}\mathcal{R}_{2\bar{O}}(\tau)+\mathcal{A}_{2O}{\rm Pr}\left(\mathcal{E}_{2OC,0}\right)\mathcal{R}_{2OC}(\tau)\notag\\
&+\mathcal{A}_{2O}\left(1-{\rm Pr}\left(\mathcal{E}_{2OC,0}\right)\right)\mathcal{R}_{2O\bar{C}}(\tau)
\end{align}
where  $\mathcal{A}_{k}\define {\rm Pr}\left(u_{0}\in\mathcal{U}_{k}\right)$ ($k\in\{1,2\bar{O},2O\}$) are given in \cite{singh13}, ${\rm Pr}\left(\mathcal{E}_{2OC,0}\right)$ is the probability that $u_{0}\in\mathcal{U}_{2O}$ is selected for IN: ${\rm Pr}\left(\mathcal{E}_{2OC,0}\right)
=\frac{U\lambda_{1}\mathcal{A}_{2}}{\lambda_{2}\mathcal{A}_{2O}}\left(1-\left(1+\frac{\lambda_{2}\mathcal{A}_{2O}}{3.5\lambda_{1}\mathcal{A}_{2}}\right)^{-3.5}\right)-\sum_{n=1}^{U}\frac{U}{n}{\rm Pr}\left(\hat{U}_{2O_{a},0}=n\right)+\sum_{n=1}^{U}{\rm Pr}\left(\hat{U}_{2O_{a},0}=n\right)$,
with $\mathcal{A}_{2}=\mathcal{A}_{2O}+\mathcal{A}_{2\bar{O}}$, ${\rm Pr}\left(\hat{U}_{2O_{a},0}=n\right)\approx\frac{3.5^{3.5}\Gamma\left(n+3.5\right)}{\Gamma(n)\Gamma(3.5)}\left(\frac{\lambda_{2}\mathcal{A}_{2O}}{\mathcal{A}_{2}\lambda_{1}}\right)^{n-1}\left(3.5+\frac{\lambda_{2}\mathcal{A}_{2O}}{\mathcal{A}_{2}\lambda_{1}}\right)^{-\left(n+3.5\right)}$ ($n\ge1$), $\mathcal{R}_{k}(\tau) \define {\rm E}_{L_{0,j_{k}}}\left[\mathcal{S}_{k}\left(f\left(\frac{L_{0,j_{k}}\tau}{W}\right)\right)\right]$ for all $k\in\mathcal{K}$, where $L_{0,j_{k}}$ is the load of the typical user's serving BS when $u_{0}\in\mathcal{U}_{k}$. Specifically, 
\begin{align}
&\mathcal{R}_{1}(\tau) = \sum_{n\ge 1}{\rm Pr}\left(L_{0,1}=n\right)\mathcal{S}_{1}\left(f\left(\frac{n\tau}{W}\right)\right)\\
&\mathcal{R}_{2\bar{O}}(\tau)=\sum_{n\ge1}{\rm Pr}\left(L_{0,2}=n\right)\mathcal{S}_{2\bar{O}}\left(f\left(\frac{n\tau}{W}\right)\right)\\
&\mathcal{R}_{2OC}(\tau)=\sum_{n\ge1}{\rm Pr}(L_{0,2}=n)\mathcal{S}_{2OC}\left(f\left(\frac{n\tau}{W}\right)\right)\\
&\mathcal{R}_{2O\bar{C}}(\tau)=\sum_{n\ge1}{\rm Pr}(L_{0,2}=n)\mathcal{S}_{2O\bar{C}}\left(f\left(\frac{n\tau}{W}\right)\right)
\end{align}
where $\mathcal{S}_{k}(\cdot)$ are given by (\ref{eq:CP1})--(\ref{eq:CP2ObarC}) at the top of the next page. 
\begin{figure*}[!t]
\begin{align}
\mathcal{S}_{1}(\beta)=& \sum_{u=0}^{U}\left(\int_{0}^{\infty}\sum_{n=0}^{N_{1}-u-1}\frac{1}{n!}\sum_{n_{1}=0}^{n}\binom{n}{n_{1}}\tilde{\mathcal{L}}_{I_{1}}^{(n_{1})}\left(s,y\right)\Big|_{s=\beta y^{\alpha_{1}}}\tilde{\mathcal{L}}_{I_{2}}^{(n-n_{1})}\left(s,\left(\frac{P_{2}B}{P_{1}}\right)^{\frac{1}{\alpha_{2}}}y^{\frac{\alpha_{1}}{\alpha_{2}}}\right)\Big|_{s=\beta y^{\alpha_{1}}\frac{\rho_{2}}{\rho_{1}}}f_{Y_{1}}(y){\rm d}y\right)\notag\\
&\times{\rm Pr}\left(u_{2OC,0}=u\right)\label{eq:CP1}\\
\mathcal{S}_{2\bar{O}}(\beta)&=\int_{0}^{\infty}\sum_{n=0}^{N_{2}-1}\frac{1}{n!}\sum_{n_{1}=0}^{n}\binom{n}{n_{1}}\tilde{\mathcal{L}}_{I_{1}}^{(n_{1})}\left(s,\left(\frac{P_{1}}{P_{2}}\right)^{\frac{1}{\alpha_{1}}}y^{\frac{\alpha_{2}}{\alpha_{1}}}\right)\Big|_{s=\beta y^{\alpha_{2}}\frac{\rho_{1}}{\rho_{2}}}\tilde{\mathcal{L}}_{I_{2}}^{(n-n_{1})}\left(s,y\right)\Big|_{s=\beta y^{\alpha_{2}}}f_{Y_{2}}(y){\rm d}y\\
\mathcal{S}_{2OC}(\beta)&=\int_{0}^{\infty}\int_{\left(\frac{P_{2}}{P_{1}}\right)^{\frac{1}{\alpha_{2}}}x^{\frac{\alpha_{1}}{\alpha_{2}}}}^{\left(\frac{BP_{2}}{P_{1}}\right)^{\frac{1}{\alpha_{2}}}x^{\frac{\alpha_{1}}{\alpha_{2}}}}\sum_{n=0}^{N_{2}-1}\frac{1}{n!}\sum_{n_{1}=0}^{n}\binom{n}{n_{1}}\tilde{\mathcal{L}}_{I_{1}}^{(n_{1})}\left(s,r_{1k}\right)\Big|_{s=\beta Y_{j_{k}}^{\alpha_{j_{k}}}\frac{\rho_{1}}{\rho_{j_{k}}}}\tilde{\mathcal{L}}_{I_{2}}^{(n-n_{1})}\left(s,r_{2k}\right)\Big|_{s=\beta Y_{j_{k}}^{\alpha_{j_{k}}}\frac{\rho_{2}}{\rho_{j_{k}}}}f_{Y_{1},Y_{2}}(x,y){\rm d}y{\rm d}x\\
\mathcal{S}_{2O\bar{C}}(\beta)&=\int_{0}^{\infty}\int_{\left(\frac{P_{2}}{P_{1}}\right)^{\frac{1}{\alpha_{2}}}x^{\frac{\alpha_{1}}{\alpha_{2}}}}^{\left(\frac{BP_{2}}{P_{1}}\right)^{\frac{1}{\alpha_{2}}}x^{\frac{\alpha_{1}}{\alpha_{2}}}}\sum_{n=0}^{N_{2}-1}\frac{1}{n!}\sum_{(q_{a})_{a=1}^{3}\in\mathcal{Q}_{3}}\binom{n}{q_{1},q_{2},q_{3}}\tilde{\mathcal{L}}_{I_{1}}^{(q_{1})}\left(s,x\right)\Big|_{s=\beta y^{\alpha_{2}}\frac{\rho_{1}}{\rho_{2}}}\tilde{\mathcal{L}}_{I_{2}}^{(q_{2})}\left(s,y\right)\Big|_{s=\beta y^{\alpha_{2}}}\Gamma\left(q_{3}+1\right)\notag\\
&\hspace{6mm}\times \left(\beta\frac{\rho_{1}y^{\alpha_{2}}}{\rho_{2}x^{\alpha_{1}}}\right)^{q_{3}}\left(1+\beta\frac{\rho_{1}y^{\alpha_{2}}}{\rho_{2}x^{\alpha_{1}}}\right)^{-\left(q_{3}+1\right)} f_{Y_{1},Y_{2}}(x,y){\rm d}y{\rm d}x\label{eq:CP2ObarC}
\end{align}
\normalsize \hrulefill
\end{figure*}
$f(x)=2^{x}-1$, ${\rm Pr}\left(L_{0,j}= n\right) = \frac{3.5^{3.5}\Gamma\left(n+3.5\right)\left(\frac{\lambda_{u}\mathcal{A}_{j}}{\lambda_{j}}\right)^{n-1}}{\Gamma(3.5)(n-1)!\left(\frac{\lambda_{u}\mathcal{A}_{j}}{\lambda_{j}}+3.5\right)^{n+3.5}}$ ($j=1,2$), 
$f_{Y_{1}}(y)$ and $f_{Y_{2}}(y)$ are given in \cite{singh13}, $f_{Y_{1},Y_{2}}(x,y)$ is given in \cite{sakr14}, $\mathcal{Q}_{3}\define\{(q_{a})_{a=1}^{3}|q_{a}\in\mathbb{N}^{0},\sum_{a=1}^{3}q_{a}=n\}$,
\begin{align}\label{eq:LTtld_diff}
\tilde{\mathcal{L}}_{I_j}^{(m)}\left(s,r_{jk}\right)=&\mathcal{L}_{I_j}\left(s,r_{jk}\right)\sum_{(p_{a})_{a=1}^{m}\in\mathcal{M}_{m}}\frac{m!}{\prod_{a=1}^{m}p_{a}!}\notag\\
&\hspace{-26mm}\times\prod_{a=1}^{m}\left(\frac{2\pi}{\alpha_{j}}\lambda_{j}s^{\frac{2}{\alpha_{j}}}B^{'}\left(1+\frac{2}{\alpha_{j}},a-\frac{2}{\alpha_{j}},\frac{1}{1+sr_{jk}^{-\alpha_{j}}}\right)\right)^{p_{a}}
\end{align}
with $\mathcal{M}_{m}\define \Big\{(p_{a})_{a=1}^{m}|p_{a}\in\mathbb{N}^{0},\sum_{a=1}^{m}a\cdot p_{a}=m\Big\}$, $\mathcal{L}_{I_j}\left(s,r_{jk}\right)=\exp\left(-\frac{2\pi \lambda_{j}}{\alpha_{j}}s^{\frac{2}{\alpha_{j}}}B^{'}\left(\frac{2}{\alpha_{j}},1-\frac{2}{\alpha_{j}},\frac{1}{1+sr_{jk}^{-\alpha_{j}}}\right)\right)$, and $B^{'}(a,b,z) = \int_{z}^{1}u^{a-1}(1-u)^{b-1}{\rm d}u$ $(0<z<1)$, 
\begin{align}
{\rm Pr}\left(u_{2OC,0}=n\right)=
\begin{cases}
&{\rm Pr}\left(U_{2O_{a},0}=n\right),\hspace{6mm} {\rm for}\; 0\le n<U\\
&\sum_{u=n}^{\infty} {\rm Pr}\left(U_{2O_{a},0}=u\right),\; {\rm for}\;\; n=U
\end{cases}
\notag
\end{align}
where ${\rm Pr}\left(U_{2O_{a},0}=n\right) \approx \frac{3.5^{3.5}\Gamma\left(n+3.5\right)}{\Gamma(3.5)n!}\left(\frac{\lambda_{2}\mathcal{A}_{2O}}{\mathcal{A}_{2}\lambda_{1}}\right)^{n}$\\$\times\left(3.5+\frac{\lambda_{2}\mathcal{A}_{2O}}{\mathcal{A}_{2}\lambda_{1}}\right)^{-\left(n+3.5\right)}$ $(n\ge 0)$.
\end{theorem}

Note that the expression in (\ref{eq:CPrate_IN_overall}) is difficult to compute and analyze due to the infinite summations over $n$. To simplify the rate coverage probability expression $\mathcal{R}(\tau)$ in (\ref{eq:CPrate_IN_overall}), we use the mean of the random load (i.e., ${\rm E}\left[L_{0,1}\right]$ or ${\rm E}\left[L_{0,2}\right]$) to approximate it (i.e., $L_{0,1}$ or $L_{0,2}$) \cite{singh13}. The simplification is thus achieved due to the elimination of the infinite summation over $n$. The rate coverage probability with mean load approximation (MLA) is given by:
\begin{corollary}\label{cor:CPrate_MLA}
The rate coverage probability with MLA is  
\begin{align}\label{eq:CPrateMLA_IN}
\mathcal{\bar{R}}(\tau)=&\mathcal{A}_{1}\mathcal{\bar{R}}_{1}(\tau)+\mathcal{A}_{2\bar{O}}\mathcal{\bar{R}}_{2\bar{O}}(\tau)+\mathcal{A}_{2O}{\rm Pr}\left(\mathcal{E}_{2OC,0}\right)\mathcal{\bar{R}}_{2OC}(\tau)\notag\\
&+\mathcal{A}_{2O}\left(1-{\rm Pr}\left(\mathcal{E}_{2OC,0}\right)\right)\mathcal{\bar{R}}_{2O\bar{C}}(\tau)
\end{align}
where $\mathcal{\bar{R}}_{1}(\tau)=\mathcal{S}_{1}\left(f\left(\frac{{\rm E}\left[L_{0,1}\right]\tau}{W}\right)\right)$, $\mathcal{\bar{R}}_{2\bar{O}}(\tau)=\mathcal{S}_{2\bar{O}}\left(f\left(\frac{{\rm E}\left[L_{0,2}\right]\tau}{W}\right)\right)$, $\mathcal{\bar{R}}_{2OC}(\tau)=\mathcal{S}_{2OC}\left(f\left(\frac{{\rm E}\left[L_{0,2}\right]\tau}{W}\right)\right)$, and $\mathcal{\bar{R}}_{2O\bar{C}}(\tau)=\mathcal{S}_{2O\bar{C}}\left(f\left(\frac{{\rm E}\left[L_{0,2}\right]\tau}{W}\right)\right)$ with $\mathcal{S}_{k}(\cdot)$ given by (\ref{eq:CP1})--(\ref{eq:CP2ObarC}). Here, ${\rm E}\left[L_{0,1}\right]=1+1.28\frac{\lambda_{u}\mathcal{A}_{1}}{\lambda_{1}}$ \cite{singh13}, and ${\rm E}\left[L_{0,2}\right]=1+1.28\frac{\lambda_{u}\mathcal{A}_{2}}{\lambda_{2}}$.
\end{corollary}

\begin{figure}[t] \centering
\includegraphics[width=0.7\columnwidth]{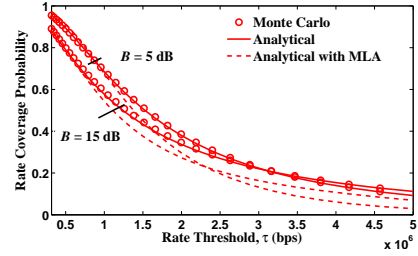}
\caption{Rate coverage probability vs. rate threshold $\tau$ for different bias factors $B$, and with $\alpha_{1}=\alpha_{2}=4$, $\frac{P_{1}}{P_{2}}=10$ dB, $N_{1}=8$, $N_{2}=4$, $U=4$, $W=10$ MHz, $\lambda_{1}=0.0001$ nodes/m$^{2}$, and $\lambda_{2}=0.0005$ nodes/m$^{2}$.}\label{fig:CPrate_diffB}
\vspace{-5mm}
\end{figure}

Fig.\ \ref{fig:CPrate_diffB} plots the rate coverage probability vs. rate threshold $\tau$ for different $B$. We see that the `Analytical' curves (i.e., $\mathcal{R}(\tau)$ in (\ref{eq:CPrate_IN_overall})) closely match with the `Monte Carlo' curves, although $\mathcal{R}(\tau)$ is derived based on the p.m.f. approximations of $U_{2O_{a},0}$ and $\hat{U}_{2O_{a},0}$. Moreover, we observe that the rate coverage probability with MLA (i.e., $\mathcal{\bar{R}}(\tau)$ in (\ref{eq:CPrateMLA_IN})) provides sufficient accuracy, especially when $\tau$ is not very large. Hence, for analytical tractability, we will investigate the rate coverage probability with MLA $\mathcal{\bar{R}}(\tau)$ in the remaining part of this paper. 


\section{Rate Coverage Probability Optimization} 
In this section, we present some properties of  the optimal design parameter which maximizes the rate coverage probability for small rate threshold regime and general rate threshold regime, respectively. 

For a fixed $B$, the optimal design parameter $U^{*}(\tau)$ is defined as follows: \footnote{We make explicit the dependence of $\mathcal{\bar{R}}(\tau)$ on $U$ in this section.}
\begin{align}\label{eq:optU_def}
U^{*}(\tau)\stackrel{\Delta}{=} {\rm arg}\; \max_{U\in\{0,1,\ldots,N_1-1\}} \mathcal{\bar{R}}(U,\tau)\;.
\end{align}
Note that (\ref{eq:optU_def}) is an integer programming problem. Moreover, the objective function $\mathcal{\bar{R}}(U,\tau)$ in (\ref{eq:CPrateMLA_IN}) is very complicated. It is thus quite challenging to obtain the closed-form optimal solution to the problem in (\ref{eq:optU_def}). 

Now, we address this challenge and characterize the optimality property of $U^{*}(\tau)$. 
Let $\Delta\mathcal{\bar{R}}(U,\tau)\define \mathcal{\bar{R}}(U,\tau)-\mathcal{\bar{R}}(U-1,\tau)$ denote the rate coverage probability change when the design parameter is changed from $U-1$ to $U$. We can show 
\begin{align}\label{eq:CPdelta_diffU}
\Delta\mathcal{\bar{R}}(U,\tau)=\mathcal{A}_{2O}\Delta\mathcal{\bar{R}}_{2O}(U,\tau)-\mathcal{A}_{1}\left|\Delta\mathcal{\bar{R}}_{1}(U,\tau)\right|
\end{align} 
where $\Delta\mathcal{\bar{R}}_{1}(U,\tau)\define\mathcal{\bar{R}}_{1}(U,\tau)-\mathcal{\bar{R}}_{1}(U-1,\tau)<0$ denotes the rate coverage probability change of a macro-user, and $\Delta\mathcal{\bar{R}}_{2O}(U,\tau)\define \mathcal{\bar{R}}_{2O}(U,\tau)-\mathcal{\bar{R}}_{2O}(U-1,\tau)>0$
denotes the rate coverage probability change of an offloaded user. Here\footnote{Here, we make explicit the dependence of ${\rm Pr}\left(\mathcal{E}_{2OC,0}\right)$ on $U$.}, $\mathcal{\bar{R}}_{2O}(U,\tau)={\rm Pr}\left(U,\mathcal{E}_{2OC,0}\right)\mathcal{\bar{R}}_{2OC}(\tau)+\left(1-{\rm Pr}\left(U,\mathcal{E}_{2OC,0}\right)\right)\mathcal{\bar{R}}_{2O\bar{C}}(\tau)$. We observe from (\ref{eq:CPdelta_diffU}) that when $U$ is increased by $1$, $\Delta\mathcal{\bar{R}}(U,\tau)$ can be decomposed into two parts, namely, the ``gain" $\mathcal{A}_{2O}\Delta\mathcal{\bar{R}}_{2O}(U,\tau)$ and the ``penalty" $\mathcal{A}_{1}\left|\Delta\mathcal{\bar{R}}_{1}(U,\tau)\right|$. Whether $\Delta\mathcal{\bar{R}}(U,\tau)$ is positive or not depends on whether the ``gain" dominates the ``penalty" or not. In the following, to maximize $\mathcal{\bar{R}}(U,\tau)$, we study the properties of $\Delta\mathcal{\bar{R}}(U,\tau)$ in (\ref{eq:CPdelta_diffU}) w.r.t.\ $U$ by comparing $\mathcal{A}_{2O}\Delta\mathcal{\bar{R}}_{2O}(U,\tau)$ and $\mathcal{A}_{1}\left|\Delta\mathcal{\bar{R}}_{1}(U,\tau)\right|$.


\subsubsection{Rate coverage probability optimization when $\tau\to0$}
We first characterize the asymptotic behavior of $\Delta\mathcal{\bar{R}}_{2O}(U,\tau)$ and $\left|\Delta\mathcal{\bar{R}}_{1}(U,\tau)\right|$ when $\tau\to0$, which is shown as follows:
\begin{lemma}\label{prop:CPloss_macro}
When $\tau\to0$, we have\footnote{$f(x)=\Theta\left(g(x)\right)$ means that $\lim_{x\to0}\frac{f(x)}{g(x)}=c$ where $0<c<\infty$.} $\Delta\mathcal{\bar{R}}_{2O}(U,\tau)=\Theta\left(\tau^{N_{2}}\right)$ and $\left|\Delta\mathcal{\bar{R}}_{1}(U,\tau)\right|=\Theta\left(\tau^{N_1-U}\right)$.
\end{lemma}


According to (\ref{eq:CPdelta_diffU}), \emph{Lemma \ref{prop:CPloss_macro}}, and noting that $\mathcal{A}_{2O}$ and $\mathcal{A}_{1}$ are independent of $\tau$, we have 
\begin{align}\label{eq:CPchange_order}
\Delta\mathcal{\bar{R}}(U,\tau)
=
\begin{cases}
\Theta\left(\tau^{N_{2}}\right)>0,\;\hspace{10.5mm}U<N_{1}-N_{2}\\
\Theta\left(\tau^{N_{2}}\right)-\Theta\left(\tau^{N_{2}}\right),\;U=N_{1}-N_{2}\\
\Theta\left(\tau^{N_{1}-U}\right)<0,\;\hspace{6mm}U>N_{1}-N_{2}\\
\end{cases}
\hspace{-3mm}.
\end{align}
Since $U^{*}(\tau)$ satisfies $\Delta\mathcal{\bar{R}}(U^{*}(\tau),\tau)>0$ and $\Delta\mathcal{\bar{R}}(U^{*}(\tau)+1,\tau)\le0$, we see from (\ref{eq:CPchange_order}) that $U^{*}(\tau)$ should be in the set $\{N_{1}-N_{2}-1,N_{1}-N_{2}\}$, and the exact value of $U^{*}(\tau)$ depends on whether $\Delta\mathcal{\bar{R}}(U,\tau)$ is positive or not when $U=N_{1}-N_{2}$ (i.e., the second case in (\ref{eq:CPchange_order})), i.e., whether the coefficient in $\Theta\left(\tau^{N_{2}}\right)$ of $\mathcal{A}_{2O}\Delta\mathcal{\bar{R}}_{2O}(U,\tau)$ (i.e., the first one) is larger than that in $\Theta\left(\tau^{N_{2}}\right)$ of $\mathcal{A}_{1}\left|\Delta\mathcal{\bar{R}}_{1}(U,\tau)\right|$ (i.e., the second one) or not. 
According to the above discussions, the optimal design parameter $U^{*}(\tau)$ is given in the following theorem: 
\begin{theorem}\label{prop:optU_lowbeta}
When $\tau\to0$, the optimal design parameter $U^{*}(\tau)\to U_{0}^{*}$, where $U_{0}^{*}\in\{N_{1}-N_{2}-1,N_{1}-N_{2}\}$.
\end{theorem}

\begin{figure}[t] \centering
\includegraphics[width=0.75\columnwidth]
{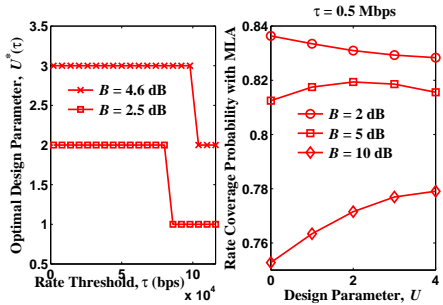}
\caption{Optimal design parameter $U^{*}(\tau)$, and with $\alpha_{1}=\alpha_{2}=3$, $\frac{P_{1}}{P_{2}}=10$ dB, $W=10\times 10^{6}$ Hz, $N_{1}=5$, $N_{2}=2$, $\lambda_{u}=0.01$ nodes/m$^{2}$, $\lambda_{1}=0.0001$ nodes/m$^{2}$, and $\lambda_{2}=0.0015$ nodes/m$^{2}$.}\label{fig:optUvstau}
\end{figure}

\emph{Theorem \ref{prop:optU_lowbeta}} shows that when $\tau\to0$, the optimal design parameter $U^{*}(\tau)$ converges to a fixed value in the set $\{N_{1}-N_{2}-1,N_{1}-N_{2}\}$, which is only related to the number of antennas at each macro-BS and each pico-BS. Fig.\ \ref{fig:optUvstau}(a) plots the optimal design parameter $U^{*}(\tau)$ vs. rate threshold $\tau$ for different $B$. We can see that $U^{*}(\tau)$ converges to a fixed value $U_{0}^{*}\in\{N_{1}-N_{2}-1,N_{1}-N_{2}\}$ when $\tau$ is sufficiently small. Specifically, we can see that $U^{*}(\tau)=N_1-N_2=3$ at $B=4.6$ dB, and $U^{*}(\tau)=N_1-N_2-1=2$ at $B=2.5$ dB. These observations verify \emph{Theorem \ref{prop:optU_lowbeta}}. In addition, we see that $U^{*}(\tau)$ is larger for a larger $B$.

\begin{figure}[t]
\centering
\subfigure[$N_{1}=8$, $N_{2}=6$, $\eta^{*}(\tau)=0.01$ at $B^{*}_{\rm ABS}$]{
\includegraphics[height=0.5\columnwidth,width=0.465\columnwidth]
{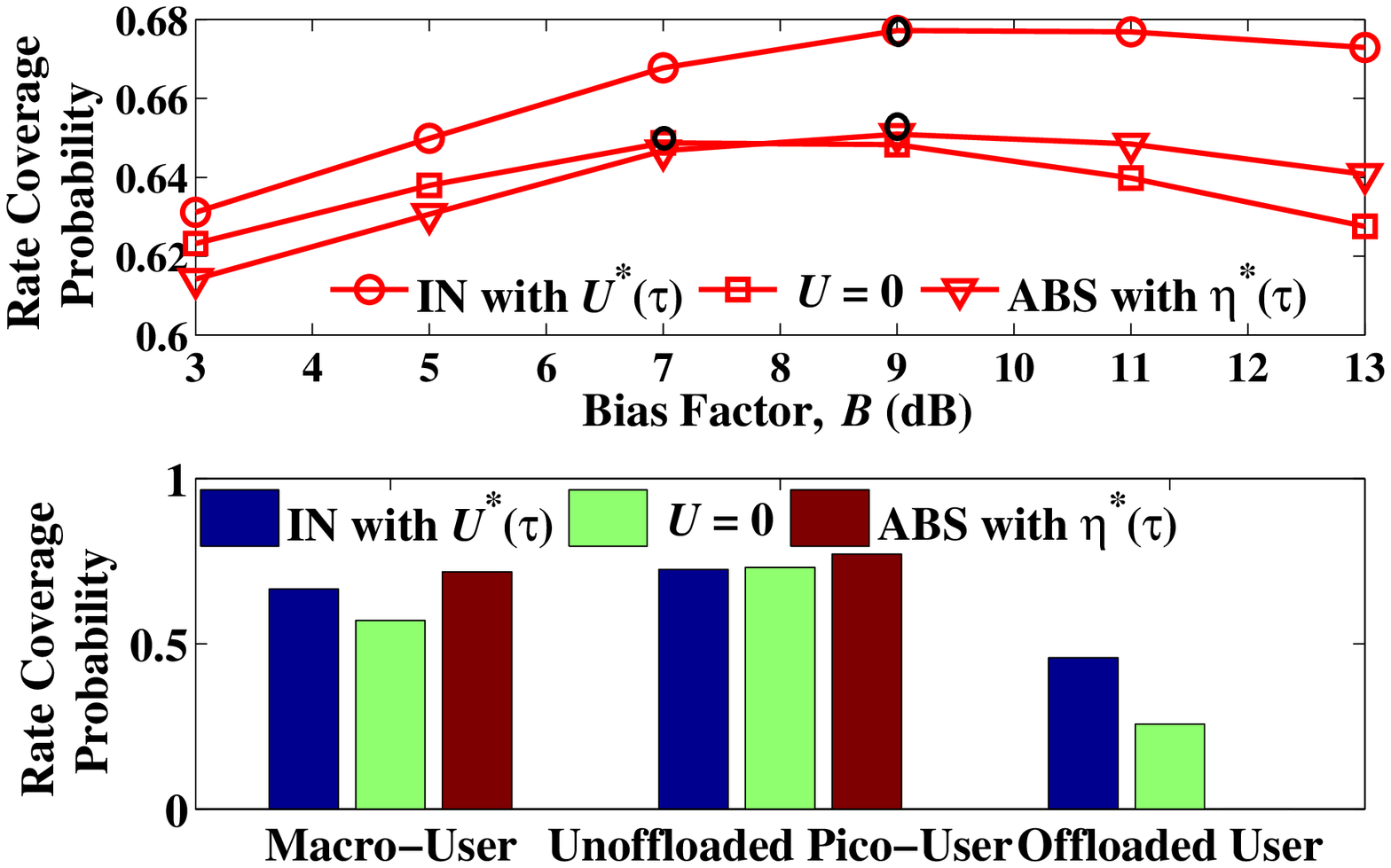}
\label{fig:N18_N26}
}
\subfigure[$N_{1}=18$, $N_{2}=16$, $\eta^{*}(\tau)=0.19$ at $B^{*}_{\rm ABS}$]{
\includegraphics[height=0.5\columnwidth,width=0.465\columnwidth]
{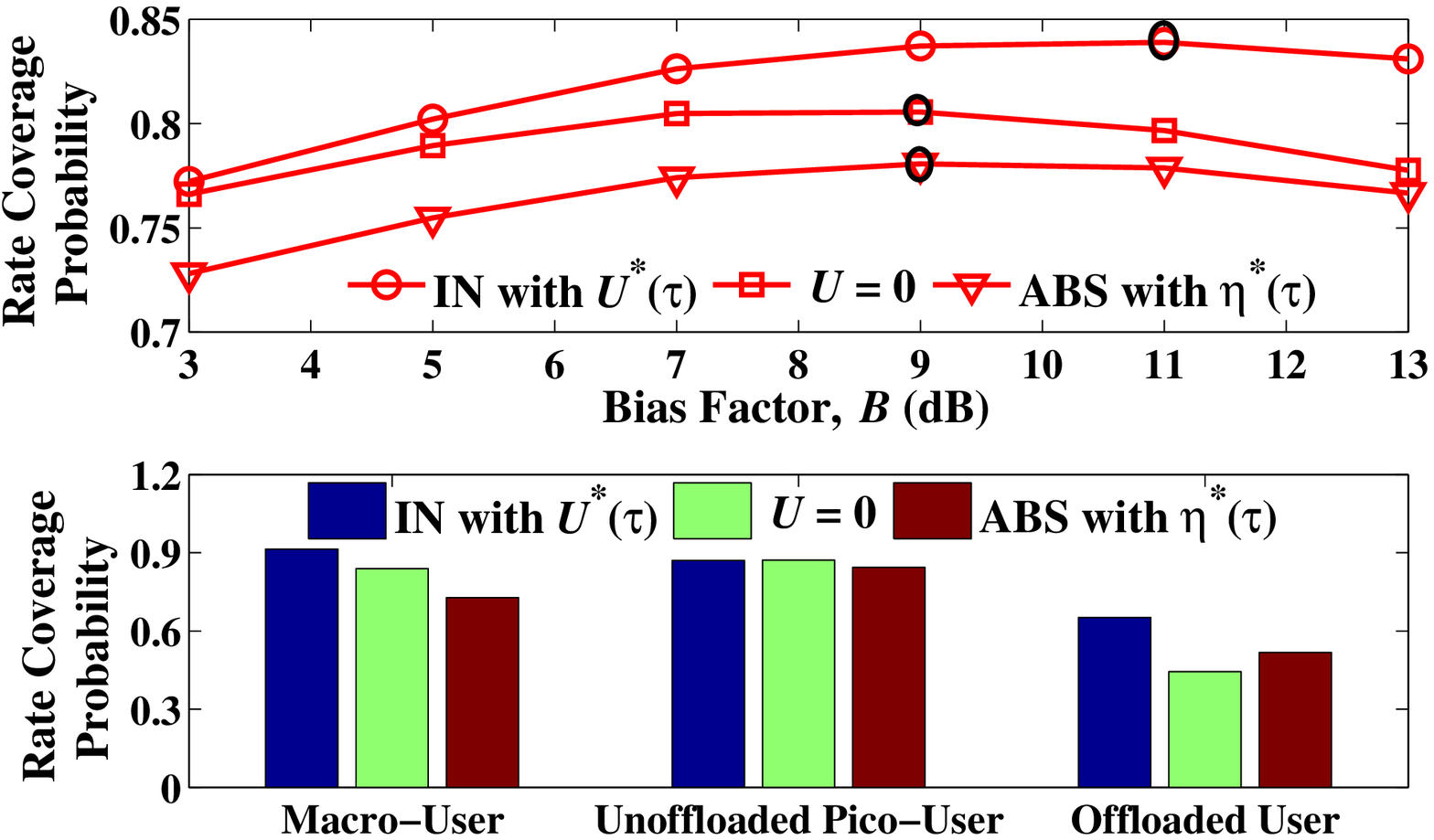}
\label{fig:N118_N216}
}
\caption{Rate coverage probability vs. bias factors $B$, and with $\alpha_{1}=4.5$, $\alpha_{2}=4.7$, $\frac{P_{1}}{P_{2}}=13$ dB, $W=10\times 10^{6}$ Hz, $\tau=5\times 10^{5}$ bps, $\lambda_{1}=0.00008$ nodes/m$^{2}$, $\lambda_{2}=0.001$ nodes/m$^{2}$, and $\lambda_{u}=0.05$ nodes/m$^{2}$. In the figures on top, the points at $B^{*}_{\rm IN}$, $B^{*}_{U=0}$, and $B^{*}_{\rm ABS}$ are highlighted using black ellipse. In the figures at the bottom, the rate coverage probability of each user type in different schemes are plotted at $B^{*}_{\rm IN}$, $B^{*}_{U=0}$, and $B^{*}_{\rm ABS}$, respectively.}\label{fig:CPratevsBdb_tau0p5MHz}
\vspace{-5mm}
\end{figure}

\subsubsection{Rate coverage probability for general $\tau$} 
Unlike the case for small $\tau$, the ``gain" $\mathcal{A}_{2O}\Delta\mathcal{\bar{R}}_{2O}(U,\tau)$ and the ``penalty" $\mathcal{A}_{1}\left|\Delta\mathcal{\bar{R}}_{1}(U,\tau)\right|$ are not determined by the order terms $\Theta\left(\tau^{N_{2}}\right)$ and $\Theta\left(\tau^{N_{1}-U}\right)$, respectively. The optimal design parameter $U^{*}(\tau)$ thus also depends on other system parameters besides $N_1$ and $N_2$. Fig.\ \ref{fig:optUvstau}(b) plots the rate coverage probability with MLA vs. $U$ for different $B$. We can see that besides $N_{1}-N_{2}-1$ and $N_{1}-N_{2}$, the optimal design parameter $U^{*}(\tau)$ can also take other values in set $\{0,1,\ldots,N_{1}-1\}$. In particular, from Fig.\ \ref{fig:optUvstau}(b), we see that $U^{*}(\tau)$ can be $0$ (at $B=2$ dB), $2$ (at $B=5$ dB), and $N_{1}-1=4$ (at $B=10$ dB). Interestingly, Fig.\ \ref{fig:optUvstau}(b) indicates that, for general $\tau$, the optimal parameter $U^{*}(\tau)$ increases with $B$, which is consistent with the case for small $\tau$ in Fig.\ \ref{fig:optUvstau}(a). 

\section{Numerical Results}
In this section, we compare the \emph{IN scheme under $U^{*}(\tau)$} with the \emph{simple offloading scheme without interference management (i.e., $U=0$)} and \emph{multi-antenna ABS under $\eta^{*}(\tau)$} (i.e., the multi-antenna version of ABS in \cite{singh13}). Specifically, $\eta\in(0,1)$ is the continuous control variable for ABS, where $\eta$ represents the resource fraction for serving the offloaded users only, while $1-\eta$ represents the resource fraction for serving the macro-users and unoffloaded pico-users simultaneously. In the simulation, the optimal control variable $\eta^{*}(\tau)$ of ABS is obtained by bisection method with $N_{1}$ iterations, while the optimal (discrete) design parameter $U^{*}(\tau)$ of the IN scheme is obtained by exhaustive search over $\{0,1,\ldots,N_{1}-1\}$.

Fig.\ \ref{fig:CPratevsBdb_tau0p5MHz} plots the rate coverage probability vs. bias factor $B$ for all the three schemes. Note that under the parameters used in the simulation, we have sufficient offloaded users, and the dominant macro-interferer is sufficiently strong compared to the remaining macro-interferers. We see from Fig.\ \ref{fig:CPratevsBdb_tau0p5MHz} that the optimal rate coverage probability (maximized over $B$) of the IN scheme is larger than those of both the simple offloading scheme and ABS. \footnote{Note that the IN scheme may not provide gains in all scenarios, as suggested in Fig.\ \ref{fig:optUvstau}.} We denote the optimal $B$ for the IN scheme, simple offloading scheme, and ABS as $B^{*}_{\rm IN}$, $B^{*}_{U=0}$, and $B^{*}_{\rm ABS}$, respectively. In particular, when $B^{*}_{\rm IN}$ is sufficiently large (i.e., sufficient offloaded users), if $N_{1}$ is sufficiently large (e.g., $N_{1}=8$), the IN scheme can achieve good performance gains over both the simple offloading scheme and ABS, as the penalty of sacrificing some DoF for IN becomes minor.  
From the figures at the bottom of Fig.\ \ref{fig:N18_N26} and Fig.\ \ref{fig:N118_N216}, we see that the rate coverage probability of the offloaded users in the IN scheme at $B^{*}_{\rm IN}$ is larger than those in the simple offloading scheme at $B^{*}_{U=0}$ and in ABS at $B^{*}_{\rm ABS}$. This is because the offloaded users in the IN scheme do not have (time or frequency) resource sacrifice, and their dominant macro-interferers can be cancelled. However, the offloaded users in ABS suffer from resource limitations, and the offloaded users in the simple offloading scheme suffer from the strong interference caused by their dominant macro-interferers.



\end{document}